\documentclass[pra,a4paper,aps,onecolumn,showpacs,superscriptaddress,groupedaddress]{revtex4}
\usepackage[dvips]{graphicx}
\usepackage{ae}
\usepackage[T1]{fontenc}
\usepackage[ansinew]{inputenc}
\usepackage{amsmath}
\usepackage{amssymb}
\usepackage{graphicx}
\usepackage{caption}
\usepackage{color}
\usepackage[colorlinks]{hyperref}
\usepackage{lscape}
\begin{document}
\title{ Arbitrary Measurement dependence in tripartite non-locality }
\author{Sk Sahadat Hossain}
\email{sk.shappm2009@gmail.com}
\affiliation{Department of  Mathematics, Nabagram Hiralal Paul College, Hooghly-721246, West Bengal, India}
\author{Mostak Kamal Molla}
\email{mostakkamal@gmail.com}
\affiliation{Department of Applied Mathematics, University of Calcutta, 92, A.P.C. Road, Kolkata-700009, India }
\author{Amit Kundu}
\email{amit8967@gmail.com}
\affiliation{Department of Applied Mathematics, University of Calcutta, 92, A.P.C. Road, Kolkata-700009, India }
\author{Biswajit Paul}
\email{biswajitpaul4@gmail.com}
\affiliation{Dept.Of Mathematics, Balagarh Bijoy Krishna Mahavidyalaya, Balagarh, Hooghly-712501, West Bengal, India}
\author{Indrani Chattopadhyay}
\email{icappmath@caluniv.ac.in}
\affiliation{Department of Applied Mathematics, University of Calcutta, 92, A.P.C. Road, Kolkata-700009, India }
\author{Debasis Sarkar}
\email{dsarkar1x@gmail.com, dsappmath@caluniv.ac.in}
\affiliation{Department of Applied Mathematics, University of Calcutta, 92, A.P.C. Road, Kolkata-700009, India }
\begin{abstract}
The assumption of measurement independence is required for a local deterministic model to conduct a Bell test. The violation of a Bell inequality by such a model implies that this assumption must be relaxed. The degree to which the assumption needs to be relaxed to achieve violation of some bipartite Bell inequalities, has been investigated recently in [Phys. Rev. Lett. 105, 250404(2010), Phys. Rev. A 99, 012121(2019)]. In this work, we study the minimum degree of relaxation required to simulate violations of various well-known tripartite Bell inequalities and opens the possibility of relaxation in multipartite scenario. Local deterministic models are also provided to achieve the violations of these Bell inequalities.
\end{abstract}
\date{\today}
\maketitle
\section{Introduction}
In quantum theory, Bell-like inequalities \citep{1,2} provide constraints on certain quantum correlations based on the realism, locality and measurement independence. The violation of these inequalities confirms the presence of non-locality in such correlations. Numerous experiments with entangled particles are performed, consistent with the predictions of quantum mechanics, leading to strong empirical support for quantum theory \citep{3,4,5,6,7,8,9,10}. These tests suggest us that at least one of the prominent reasonable assumptions ( realism, locality) necessary to derive Bell's inequality must fail to hold in the physical world. Meanwhile, if one or more assumptions required to generate Bell inequalities are relaxed, it brings up "loopholes" whereby local hidden variable models could explain all previous Bell-violating experiments \citep{11,12,13}. Thus, it is important to address as many as possible loopholes in a single experimental test \citep{3,4,5,6,7,8,14}.

Another crucial assumption in the formulation of Bell inequalities is "measurement independence" where the observers have complete freedom to select detector settings in an experimental test. Relaxation of this assumption leads to "measurement independence" or "freedom of choice" loophole. Recently, this issue has been received much attention to the researcher and recent theoretical developments suggest us that the use of Bell tests to exclude local hidden variable description of quantum mechanics is most vulnerable to this individual loophole \citep{15,16,17,18,19,20,21}.

This work is motivated by the current theoretical developments that relax the measurement independence assumption \citep{15,16,17,18,19,20,21,22,23,24,25}, and the experiments that constrain such models \citep{9,14,26,27,28}. Several assumptions of this empirical aspects of quantum non-locality has significant practical relevance for many entanglement based technologies, such as, device-independent quantum key distributions \citep{29,30,31,32}, random number generation and randomness expansions \citep{33,34,35,36,37,38}, are few among them.

Interestingly, we know that in bipartite system  some non-quantum resources are useful for simulating singlet state correlations by suitable relaxation of the said assumptions of measurement independence \citep{17,18,19,21}, while obeying local causality. It was shown in \citep{17}, a deterministic, no signaling model can simulate the singlet state correlations and maximum Bell violation could be obtained by sacrifice of 14$ \% $ and 33$ \% $ measurement freedom respectively. In such scenario, measurement dependence \citep{17,18,39} was defined by the parameters $ M_{1} $ and $ M_{2} $ for different parties  ( namely, Alice and Bob) respectively. While for locally casual models that entertain measurement independence, must immolate 100$ \% $ of determinism or locality to simulate singlet state correlations \citep{15,17,18,40}. Friedman et al. \citep{21}, along with the above relaxation, have addressed the joint measurement relaxation (i.e., Alice and Bob both have restricted measurement freedoms), noted as $ M_{12} $. In such cases, they have developed respective deterministic, no signaling models justifying their claims. Another important work done by Toner and Bacon in \citep{40}, where they have founded that quantum correlation of a singlet state can be generated by one bit of communication (signaling correlation) where the measurement outcome is deterministic.

While the preceding works have assumed identical relaxation of measurement independence for all parties in a bipartite system \citep{17,18,21} or 100$ \% $ measurement independence for one observer and some  nonzero measurement dependence for the other observer \citep{16,19,21} in a bipartite scenario. Using these concepts, we have address the question: for a deterministic no signaling model, how much measurement dependence is needed in tripartite non-local scenario to simulate Greenberger-Horne-Zeilinger (GHZ) state correlation in both one party and two parties (bipartite) measurement settings. Initially, we have assumed relaxed measurement independence for one party measurement settings, and other two parties remains free and secondly considered measurement dependence for two parties measurement settings in genuine and standard tripartite non-local scenario. Our study based on  well known tripartite inequalities, namely, Svetlichny inequality \citep{41} (genuine tripartite non-locality), Mermin inequality \citep{42} (standard non-locality) and the NS$ _{2} $ inequality developed by Bancal et al., \citep{43}. Our motivation stems, in part, from recent attempt to address the detection loophole and measurement independent loophole experimentally \citep{44,45}, on the contrary it has significant practical relevance in quantum communication \citep{46}, device-independent quantum key distributions \citep{35,47,48} and quantum cryptography protocols \citep{49}. Although any presumption of possible measurement relaxation for either of these techniques would be highly model dependent, it is plausible that they would be capable to different amount of measurement dependence. Future test of tripartite non-locality, in which observer select distinct methods to detect detectors measurement settings, would then generically fall into the general class of our present work.

We have considered a scenario consisting of three spatially separated observers, viz., Alice, Bob and Charlie, each with two dichotomic measurement settings, where we define the amount of measurement dependence for each observer by using the parameters $ M_{1} ,  M_{2} $ and $ M_{3} $ introduced by Hall in \citep{17}, and also we introduce the two party joint measurement dependence as $ M_{12} $ (Alice-Bob), $ M_{23} $ (Bob-Charlie) and $ M_{13} $ (Alice-Charlie), respectively. In such scenario, we derive upper bounds on the above mentioned tripartite inequalities \citep{41,42,43}, for models that relax measurement independence but retain local causality. Our findings results in 41$ \% $ measurement dependence for Svetlichny inequality and NS$ _{2} $ inequality to reproduce GHZ state correlation for a deterministic no signaling models, where as, for Mermin inequality this measurement dependence shows asymmetric character in one-sided and two-sided measurements scenario. We have provided deterministic no signaling models to simulate such relaxed outcomes. By contrast, to simulate general no-signaling bounds for Svetlichny non-locality and NS$ _{2} $ non-locality the respective party (parties) sacrifices its full freedom of measurement choice in a deterministic measurement dependence scenario.  

Our paper is organized as follows: In Section II, we review some of the basic tools of tripartite non-locality,  and measurement dependence \citep{17,18}. In Section III, we derive the relaxed inequalities for one-sided measurement settings scenario for Mermin inequality, Svethlichny inequality and $ NS_{2} $ inequality. In Section IV, we presented modified Mermin inequality and Svetlichny inequality under bi party relaxed measurements scenario. Section V ended with conclusion. In Appendices A, B and C we present several steps of one sided relaxed Mermin inequality, Svetlichny inequality and NS$ _{2} $ inequality that are required in Sec. III, while we illustrate some steps of the models presented in Sec. IV as Appendices D and E respectively.

\section{ Basic tools } \label{md}
\subsection{Tripartite Nonlocality}\label{secI}
Here we have recapitulate some of the basic ideas of tripartite non-locality, necessary for our study. Consider a tripartite system consisting of three spatially separated parties, say, Alice, Bob and Charlie, each performs two dichotomic measurements on their respective sub-systems. Suppose, $ x,\; x^{'}$ be the inputs and $ a,\; a^{'}$ be the outputs of Alice's side, similarly $ y,\; y^{'}$ and $ b,\; b^{'}$ be that of Bob's and $ z,\; z^{'}$ and $ c,\; c^{'} $ for Charlie's, where $ x,\;y,\;z,\;x^{'},\;y^{'},\;z^{'}$ $\in $ $\lbrace 0,1\rbrace $,  and $ a,\;b,\;c,\;a^{'},\;b^{'},\;c^{'} $  $ \in $ $\lbrace 1,-1 \rbrace $. Corresponding correlations are described by the joint probability distributions $ \lbrace P (a,\;b,\;c\vert x,\;y,\;z) \rbrace $. These correlations are said to be local if there exists a local hidden variable model ($ \mu $) such that the correlations can be express as,
\begin{equation}\label{e1}
P(a,\;b,\;c\vert x,\;y,\;z) = \int\rho(\mu) \;p_{\mu}(a\vert x)\;p_{\mu}(b\vert y)\;p_{\mu}(c\vert z)\; d\mu  
\end{equation}
where $ \int \rho(\mu)\; d\mu = 1 $.

For discrete systems, we have the relation,
\begin{equation}\label{e2}
P(a,\;b,\;c\vert x,\;y,\;z) = \Sigma_{\mu}  q_{\mu}  \;p_{\mu}(a\vert x)\; p_{\mu}(b\vert y)\;p_{\mu}(c\vert z),
\end{equation}
for all $x,\; y,\; z,\; a,\; b,\; c $ and the probability distribution $ q_{\mu} $ respecting the relations, $ 0\leq q_{\mu}\leq 1$, $ \Sigma_{\mu} \;q_{\mu}=1 $.

The quantity $ p_{\mu}(a\vert x) $ is the conditional probability of  outcome $a$ when Alice performs the measurement $x$ on her sub-system and $ \mu $ is the hidden variable, similarly, $ p_{\mu}(b\vert y)  $, $ p_{\mu}(c \vert z) $ are defined for Bob and Charlie respectively. If the correlations cannot be written in the above form  Eq. \eqref{e1} or Eq.
 \eqref{e2} then we call them as non-local \citep{41,42,43}.
 
The standard tripartite non-locality is certified by the violation of Mermin inequality \citep{42}, which has the following expression, 
\begin{equation}\label{e3}
\vert \langle A_{1}B_{0}C_{0}\rangle +\langle A_{0}B_{1}C_{0}\rangle+\langle A_{0}B_{0}C_{1}\rangle-\langle A_{1}B_{1}C_{1}\rangle \vert \leq 2.
\end{equation}

Here$ \langle A_{x}B_{y}C_{z}\rangle=\Sigma_{a,\;b,\;c}\; abc P(a,\;b,\;c\vert x,\;y,\;z)$.
Tripartite entangled states which violate the Mermin inequality are often referred to as standard tripartite non-local states. The GHZ state, $ \vert GHZ \rangle=\frac{1}{\sqrt{2}} (\vert 000\rangle +\vert111\rangle ) $ violates the inequality upto its maximum quantum bound four \citep{49}.

However, Svetlichny \cite{41} showed that there exist tripartite correlations which can not be written as the following hybrid local-nonlocal correlation,
\begin{equation}\label{e4}
P(a,\;b,\;c\vert x,\;y,\;z) = \Sigma _{\mu} q_{\mu} P_{\mu}(a,\;b\vert x,\;y)\;P_{\mu}(c\vert z)\\
 +\Sigma _{\nu} q_{\nu} P_{\nu}(a,\;c\vert x,\;z)\;P_{\nu}(b\vert y)\\
 +\Sigma _{\xi} q_{\xi} P_{\xi}(b,\;c\vert y,\;z)\;P_{\xi}(a\vert x),
\end{equation}
where $ q_{\mu},\;q_{\nu},\;q_{\xi} $ are three probability distributions over the hidden variables $ \mu,\;\nu,\;\xi$ satisfying the relations $ 0\leq q_{\mu},\;q_{\nu},\;q_{\xi} \leq 1$ and $ \Sigma _{\mu} q_{\mu} +\Sigma_{\nu} q_{\nu} +\Sigma_{\xi} q_{\xi} = 1 $.

The above form of tripartite correlations is not fully local as in Eq. \eqref{e1} or Eq. \eqref{e2}. In this relation, non-local correlations are present between any two parties that are locally correlated with the third party (they can change their order in random experiments). If the correlations cannot be written as in Eq. \eqref{e4}, they are called genuine tripartite non-local. In \citep{43}, this type of non-locality is termed as Svetlichny non-locality \citep{41}, which has the following form,\\
\begin{equation}\label{e5}
 \vert \langle A_{0}B_{0}C_{0} \rangle+\langle A_{1}B_{0}C_{0} \rangle + \langle A_{0}B_{1}C_{0} \rangle -\langle A_{1}B_{1}C_{0} \rangle +\langle A_{0}B_{0}C_{1} \rangle -\langle A_{1}B_{0}C_{1} \rangle -\langle A_{0}B_{1}C_{1} \rangle -\langle A_{1}B_{1}C_{1} \rangle \vert \leq 4 . \\
 \end{equation}
 
The $ \vert GHZ \rangle $  and $\vert W \rangle $ states violate the inequality. Svethlichny inequality of tripartite system is such that no restriction is imposed on the bipartite correlations present in Eq. \eqref{e4}, consequently there may arise correlations which cause one-way or both-way signaling. For this reason, Bancal et al. \cite{43}, introduced more physical definition of tripartite non-locality in terms of no-signaling principle, where the correlations among observers satisfy no-signaling restrictions. For the tripartite correlations $ P(a,\;b,\;c\vert x,\;y,\;z) $   in Eq. \eqref{e4}, must satisfy the no-signaling conditions as mentioned below:
\begin{eqnarray}\label{e6}
 p_{\mu}(a\vert x) = \Sigma_{b} p_{\mu} (a,\;b\vert x,\;y); \forall  x,\;y,\;a.\\
  p_{\mu}(b\vert y) = \Sigma_{a} p_{\mu} (a,\;b\vert x,\;y); \forall  x,\;y,\;b. 
\end{eqnarray}
and similar relations hold for other bipartite correlations in Eq. \eqref{e4}. The above form of correlations are known as $ NS_{2}$ local correlations, otherwise, they are $ NS_{2} $ non-local (iff they are not fully local). In \cite{43} Bancal et al., have generated 185 Bell-type inequalities which constitute the full class of facets of $ NS_{2} $ polytope. Here for our purpose we have considered only the 99th inequality of this set, which has the following representation, \\
\begin{equation}\label{e7}
\vert \langle A_{0}B_{0} \rangle +\langle A_{0}C_{0}\rangle +\langle B_{0}C_{1}\rangle -\langle A_{1}B_{1}C_{0}\rangle +\langle A_{1}B_{1}C_{1}\rangle \vert \leq 3. 
\end{equation}

There exist correlations in tripartite system which violate Eq. \eqref{e7} but are Svetlichny local \citep{43}.

\subsection{ Quantifying  Measurement Dependence }\label{secII}
Consider an underlying hidden variable model $ \lambda $ (apparently multicomponent) such that the joint probability distribution is of the following form,  \\
\begin{equation}\label{e8}
 P(a,\;b,\;c\vert x,\;y,\;z) = \int \rho(\lambda\vert x,\;y,\;z)\; p(a,\; b,\; c\vert x,\;y,\;z,\;\lambda)\;  d\lambda,
 \end{equation}
where $ \int \rho(\lambda\vert x,\;y,\;z) d\lambda =1 $. 

Thus the hidden variable $ \lambda $ includes among its components any hidden variables that produce the correlations. The measurement independence is the property that the choice of measurement setting $x,\; y,\; z$ is independent of the underlying variable $ \lambda $ which affect the measurement outcomes, i.e.,\\
\begin{equation}\label{e9}
 \rho(\lambda\vert x,\;y,\;z) = \rho(\lambda\vert x^{'},\; y^{'},\;z^{'}) = \rho(\lambda\vert x^{'},\;y^{'},\;z),
 \end{equation}
  including all possible variations of measurement settings. For any set of joint measurement settings $(x,\; y,\; z)$,  $ (x^{'},\;y^{'},\;z^{'}),\;(x^{'},\;y^{'},\;z)  $ and other set of possible inputs, it is equivalently quantified as $ \rho (\lambda\vert x,\;y,\;z) = \rho(\lambda) $. \\
Respecting the assumptions on local causality and determinism ( throughout our present study), relaxing the assumption of measurement independence, and following the framework introduced in \citep{17,18,19,20,21}, we quantify the degree of relaxation as described below. The overall degree of measurement dependence $M$ for an observer is defined as the variation of distances between the distributions of the shared random variable for any set of joint measurement settings; i.e.,
\begin{equation}\label{e10}
M=\sup_{x,\;x^{'},\;y,\;y^{'},\;z,\;z^{'}} \int d\lambda\;\vert\rho(\lambda\vert x,\;y,\;z) -\rho(\lambda\vert x^{'},\;y^{'},\;z^{'})\vert. 
\end{equation}

The measurement dependence corresponds to a non-zero value of this quantity $ M $, for any set of joint measurement setting for Alice, Bob and Charlie. For all possible settings, $M$ will vary within $ [0,\;2]$.

The overall degree of freedom of choice $ F $ is quantified via \cite{17,18,50},
\begin{equation}\label{e11}
F\equiv 1-\frac{M}{2}. 
\end{equation}

Consequently, $ M= 2 $, corresponds to the the case in which two normalized probability distributions $ \rho ( \lambda \vert x,\;y,\;z) $ and $ \rho(\lambda\vert x^{'},\;y^{'},\;z^{'}) $ have no overlapping support for any value of $ \lambda $. Hence $F = 0$. Likewise, $ M = 0$ corresponds to the case $ F = 1 $, i.e., complete freedom of measurement choices.

Similarly, we may quantify one-sided degree of measurement dependence $ M_{1},\; M_{2} $ and $ M_{3} $, for measurement settings of Alice, Bob and Charlie respectively, as follows;
 \begin{equation}\label{e12}
M_{1} = \sup _{x,\;x^{'},\;y,\;z} \int d\lambda\; \vert\rho(\lambda \vert x,\;y,\;z)-\rho(\lambda\vert x^{'},\;y,\;z)\vert, 
 \end{equation}
 \begin{equation}\label{e13}
 M_{2} = \sup_{x,\;y,\;y^{'},\;z} \int d\lambda\; \vert \rho(\lambda\vert x,\;y,\;z)-\rho(\lambda\vert x,\;y^{'},\;z)\vert, 
 \end{equation}
 \begin{equation}\label{e14}
 M_{3} = \sup_{x,\;y,\;z,\;z^{'}} \int d\lambda\; \vert \rho(\lambda\vert x,\;y,\;z)-\rho(\lambda\vert x,\;y,\;z^{'})\vert.  
\end{equation} 

Here, the one-sided measure $ M_{1} $ quantifies the degree of measurement dependence corresponding to the variations of Alice's measurement settings, where settings of Bob and Charlie remain free. Similar interpretations hold for $ M_{2} $ and $ M_{3} $  corresponding to settings of Bob and Charlie respectively.

Applying the same arguments as above, the degree of measurement dependence for two-party (bipartite) measurement settings are quantified via \citep{17,18,21,50},
 \begin{equation}\label{e15}
M_{12} =\sup_{x,\;x^{'},\;y,\;y^{'},\;z} \left\lbrace \int d\lambda \;\vert \rho(\lambda\vert x,\;y,\;z)-\rho(\lambda\vert x^{'},\;y^{'},\;z)\vert, \int d\lambda \;\vert \rho(\lambda\vert x^{'},\;y,\;z)-\rho(\lambda\vert x,\;y^{'},\;z)\vert\right\rbrace, 
\end{equation}
\begin{equation}
M_{23} =\sup_{x,\;y,\;y^{'},\;z,\;z^{'}} \left\lbrace \int d\lambda\; \vert \rho(\lambda\vert x,\;y,\;z)-\rho(\lambda\vert x,\;y^{'},\;z^{'})\vert, \int d\lambda \;\vert \rho(\lambda\vert x,\;y,\;z^{'})-\rho(\lambda\vert x,\;y^{'},\;z)\vert\right\rbrace, 
\end{equation}
\begin{equation}
M_{13} =\sup_{x,\;x^{'},\;y,\;z,\;z^{'}} \left\lbrace \int d\lambda\; \vert \rho(\lambda\vert x,\;y,\;z)-\rho(\lambda\vert x^{'},\;y,\;z^{'})\vert, \int d\lambda\; \vert \rho(\lambda\vert x^{'},\;y,\;z)-\rho(\lambda\vert x,\;y,\;z^{'})\vert\right\rbrace. 
\end{equation}

where $ M_{12} $ quantifies the degree of measurement dependence for Alice-Bob measurement settings, while measurement settings of Charlie remains free. The rest two are quantified for measurement settings of Bob-Charlie and Alice-Charlie respectively. The degrees of freedom of choices are defined by the quantities $ F_{12} $, $ F_{23} $ and $ F_{13} $ accordingly.

\section{One-sided relaxed measurement in  tripartite system}
After a brief introduction of the basic ideas required for our present work, we now move forward to find the effect of arbitrary measurement dependence in one-sided measurement settings using Mermin inequality \citep{42}, Svetlichny inequality \citep{41} and NS$ _{2} $ inequality \citep{43}, respectively.

\subsection{Relaxed Mermin Inequality}\label{s1}
In a tripartite system with two measurement settings for each individual parties and measurements with two possible outcomes (as mentioned in Section \eqref{secI}), one can detect standard tripartite non-locality via the violation of Mermin inequality expressed below, 
 \begin{equation}\label{m1}
R= \vert\langle A_1B_0C_0 \rangle+\langle A_0B_1C_0 \rangle+\langle A_0B_0C_1 \rangle-\langle A_1B_1C_1 \rangle\vert\leq2.
\end{equation}

We assume, $ S = \langle$$A_1$$B_0$$C_0$$\rangle$+$\langle$$A_0$$B_1$$C_0$$\rangle$+$\langle$$A_0$$B_0$$C_1$$\rangle$-$\langle$$A_1$$B_1$$C_1$$\rangle$. 

Here, in a deterministic no-signaling model, we apply arbitrary measurement dependence in either of the parties measurement settings. In this model, the measurement outcomes for Alice are denoted as  $ u(A_{x},\;\lambda),\;u(A_{x'},\;\lambda)  ,$ each taking values from the set  $  \lbrace1, -1\rbrace $. Identical interpretation follows for  $ v(B_{y},\;\lambda),\;v(B_{y'},\;\lambda) $ (Bob's) and  $ w(C_{z},\;\lambda),\;w(C_{z'},\;\lambda) $ (Charlie's), respectively.

The correlation term is defined as  $\langle$$A_1$$B_0$$C_0$$\rangle$ = $\int{\rho(\lambda|x^{'},\;y,\;z)\;u(A_{x'},\;\lambda)\; v(B_{y},\;\lambda)\; w(C_{z},\;\lambda)\;d\lambda}$, and similarly other terms of the expression above. Using these quantities, the parameter $S$ is modified as (see Appendix A \eqref{sh1} for detail calculations),\\
  
\begin{equation}
\vert S \vert\leq 2+  Min  \lbrace 2,\; 2M_{1}+M_{2},\;2M_{1}+M_{3},\;2M_{2}+M_{3},\;2M_{2}+M_{1},\;2M_{3}+M_{2},\;2M_{3}+M_{1} \rbrace.
\end{equation}

For the GHZ state, $ \vert GHZ \rangle = \frac{1}{\sqrt{2}}  (\vert 000 \rangle +\vert 111 \rangle )$, it is known that the maximum quantum value of Mermin inequality for specific set of measurements is 4 \cite{49}, which is also the maximum possible bound for Mermin inequality. Hence for this local deterministic model, we assume $ 2+ 2M_{1}+M_{2} = 4 $, which implies
\begin{equation}\label{mm3}
2M_{1}+M_{2} = 2.
\end{equation}
 
Here, $  M_{1} = 0,\;M_{2} = 2 $ implies $F_{2} = 0 $, which exhibit no experimental free will for Bob's setting, i.e., in this scenario the local model sacrifices its full freedom of choice for Bob to simulate the quantum  bound of GHZ state, while measurement settings of Alice and Charlie remain free. 

Again, $ M_{2} = 0,\;M_{1} = 1 \Rightarrow F_{1} = .50 $, i.e., with 50$ \% $ measurement independence of Alice's settings, this local model can generate the GHZ state correlations, while measurement settings of Bob and Charlie remain free.

Thirdly, when we consider both the parameters nonzero, i.e., $ M_{2} = 1,\;M_{1} = .50,$ it follows that $F_{1} = 0.75,\;F_{2} = .50$. Consequently, this model can simulate the GHZ state correlations sacrificing 25$ \% $ and 50$ \% $ measurement freedom for Alice and Bob respectively.

With a similar procedure, we can find the values of $  F_{1},\;F_{2},\;F_{3} $ by analyzing other terms of this modified Mermin inequality. In such scenario, we have found that at least 50$ \% $ measurement dependence is necessary in one-sided measurement settings to achieve the quantum bound of GHZ state.

Now, we present a model which could explain this no-signaling violation of Mermin inequality,  

\begin{table} [h]
\caption{Deterministic no-signaling model I}
\begin{tabular}{c c c c c c c c c c c c }
\hline
$\lambda$ & $u(A_x,\lambda)$&$u(A_{x'},\lambda)$&$v(B_y,\lambda)$&$v(B_{y'},\lambda)$& $w(C_z,\lambda)$& $w(C_{z'},\lambda)$&$\rho(\lambda|x'yz)$ &$\rho(\lambda|xy'z)$&$\rho(\lambda|xyz')$&$\rho(\lambda|x'y'z')$\\ [1ex]
\hline
$\lambda_1$ & $a$ &$a$&$a$&$a$&$a'$&$a'$&$ 1-p_{1} $&$ 1-p_{2} $&$ 1-p_{1}-p_{2} $&$ 1-p_{1}-p_{2} $\\
$\lambda_2$ & $b$ &$b$&$b$&$b$&$b'$&$-b'$&$ p_{1} $&0&0&$p_{2}$\\
$\lambda_3$ & $c$ &$c$&$c$&$c$&$c'$&$c'$&0&$ p_{2} $&$p_{1}$&$0$\\
$\lambda_4$ & $d$ &$d$&$d$&$-d$&$d'$&$d'$&0&0&$ p_{2} $&$p_{1}$\\[1ex]
\hline
\end{tabular}
\end{table}

In the above table I, we have analyzed the GHZ state model compatible with the above modified bound of Eq. \eqref{mm3}, which contains four underlying variables $ \lambda_{1},\;\lambda_{2},\;\lambda_{3},\;\lambda_{4} $. The outcomes are specified via $ a,\;b,\;c,\;d $ $\in$ $ \lbrace 1,-1 \rbrace$, whereas we have assumed $ a^{2} = a',\;b^{2} = b',\;c^{2} = c',\;d^{2} = d' $. Here, the probability distributions are defined by the parameters $ p_{1},\;p_{2} $, with $ 0\leq $ $p_{1},\;p_{2}$ $\leq 1 $. From the table, it follows that $ S=2+2p_{1}+2p_{2} $.

Therefore comparing with the Eq. \eqref{mm3}, we have $ M_{1} = p_{1}$ and $M_{2} = 2p_{2}$, provided $ Max  \lbrace 2M_{1}+M_{2}\rbrace \leq 2 $, for the above no signaling deterministic model. Hence $ F_{1} = 1-\frac{M_{1}}{2}$ = 1 $-$ $\frac{p_{1}}{2}$, and $ F_{2} = 1-p_{2} $.

Thus this model can generate the  quantum bound of GHZ state for $ p_{1} = 1,\; p_{2} = 0; $ or $ p_{1} = 0,\; p_{2} = 1 $. So, in the first scenario a local deterministic model can generate the GHZ state bound with  50$\% $ measurement dependence for Alice's measurement settings while Bob and Charlie's measurement settings remain free. In the second site with the cost of Bob's 100$\% $ measurement freedom, we can simulate the GHZ state bound whereas Alice and Charlie are free to choose their respective measurement settings.

\subsection{Relaxed Svetlichny Inequality }
After analyzing one-sided relaxed Mermin inequality in the last subsection, to understand how it behaves with other inequalities, here we establish the effect of  measurement relaxation in one-sided measurement settings for Svetlichny inequality, which has the following representation,
\begin{equation}\label{c1}
R = \vert \langle A_0B_0C_0 \rangle +\langle A_0B_0C_1 \rangle+\langle A_1B_0C_0 \rangle-\langle A_1B_0C_1\rangle +\langle A_0 B_1C_0 \rangle - \langle A_0 B_1C_1 \rangle-\langle      A_1B_1C_0 \rangle-\langle A_1B_1C_1 \rangle \vert \leq 4. 
\end{equation}

Choose, $ S = \langle$$A_0$$B_0$$C_0$$\rangle$+$\langle$$A_0$$B_0$$C_1$$\rangle$$+$$\langle$$A_1$$B_0$$C_0$$\rangle$$-$$\langle$$A_1$$B_0$$C_1$$\rangle$+$\langle$$A_0$$B_1$$C_0$$\rangle$$-$$\langle$$A_0$$B_1$$C_1$$\rangle$$-$$\langle$$A_1$$B_1$$C_0$$\rangle$$-$$\langle$$A_1$$B_1$$C_1$$\rangle$,\\ 
where, $\langle$ $A_0$$B_1$$C_0$$\rangle$ = $\int{\rho(\lambda|x,\;y',\;z)\;u(A_x,\;B_{y'},\;\lambda)\;v(C_z,\;\lambda)\;d\lambda}$, with $-1\leq u(A_x,\;B_{y'},\;\lambda)\leq 1;\; -1\leq v(C_z,\;\lambda) \leq 1$.
Similarly, we define other terms of $ S$.

Applying one sided measurement dependence in this scenario, we have (for detail calculations, see Appendix B \eqref{sh2}),
\begin{equation}\label{mm4}
\vert S\vert \leq 4+ Min\lbrace 4,\; 2M_{1},\;2M_{2},\;2M_{3} \rbrace.
\end{equation}

We know that for the GHZ state, the maximum quantum bound of Svetlichny inequality is \cite{51} $ S=4\sqrt{2} $. For this deterministic no-signaling model we consider, $ 4+2M_{1}= 4\sqrt{2}$. Therefore, $ M_{1} = 2(\sqrt{2}-1),\; \Rightarrow F_{1} = 0.59 $.

Thus, this model demonstrate quantum violation of GHZ state by sacrificing $ 41 \% $ measurement freedom.     

For simulating general no-signaling bound \citep{53} of Svetlichny non-locality by a no-signaling deterministic model, we need measurement dependence
 $ 2M_{1} = 4 \; \Rightarrow M_{1} = 2 $, $ \Rightarrow F_{1}=0 $.
 
So, in this case, there is no experimental free will for Alice. Since the modified inequality is symmetric with respect to the parties, therefore $ M_{2}$ and $ F_{2} $ or $ M_{3}$ and $F_{3} $ will produce identical outcomes.

Now, we provide a model to establish the above results; 
\begin{table}  [h]
\caption{ Deterministic no-signaling model II}
\begin{tabular}{c c c c c c c c c c c c c c c}
\hline
$\lambda$ & $u_{1}$&$u_{2}$&$u_{3}$&$u_{4}$& $v(C_z,\lambda)$& $v(C_{z'},\lambda)$&$\rho(\lambda|xyz)$ &$\rho(\lambda|xyz')$&$\rho(\lambda|xy'z)$&$\rho(\lambda|xy'z')$&$\rho(\lambda|x'yz)$&$\rho(\lambda|x'yz')$&$\rho(\lambda|x'y'z)$&$\rho(\lambda|x'y'z')$\\ [1ex]
\hline
$\lambda_1$ &$ a$ &$a$&$a$&$-a$&$a$&$a$&$ 1-p $&$1-p$&$1-p$&$1-p$&$1-p$&$1-p$&$1-p$&$1-p$\\
$\lambda_2$ &$ b$ &$b$&$b$&$b$&$b$&$b$&$ p $&$0$&$0$&$0$&$0$&$0$&$0$&$0$\\
$\lambda_3$ &$ c$ &$c$&$c$&$c$&$c$&$c$&0&$p$&$0$&$0$&$0$&$0$&$0$&$0$\\
$\lambda_4$ &$ d$ &$d$&$d$&$d$&$d$&$d$&0&$0$&$p$&$0$&$0$&$0$&$0$&$0$\\
$\lambda_5$ &$ e$ &$-e$&$e$&$e$&$e$&$e$&0&$0$&$0$&$p$&$0$&$0$&$0$&$0$\\
$\lambda_6$ &$ f$ &$f$&$f$&$f$&$f$&$f$&0&$0$&$0$&$0$&$p$&$0$&$0$&$0$\\
$\lambda_7$ &$ g$ &$g$&$-g$&$g$&$g$&$g$&0&$0$&$0$&$0$&$0$&$p$&$0$&$0$\\
$\lambda_8$ &$ h$ &$h$&$h$&$-h$&$h$&$h$&0&$0$&$0$&$0$&$0$&$0$&$p$&$0$\\
$\lambda_9$ &$ i$ &$i$&$i$&$i$&$i$&$-i$&0&$0$&$0$&$0$&$0$&$0$&$0$&$p$\\[1ex]
\hline
\end{tabular}
\end{table}

Here $ u_{1}= u(A_x,\;B_y,\;\lambda),\,u_{2}= u(A_x,\;B_{y'},\;\lambda),\:u_{3}= u(A_{x'},\;B_y,\;\lambda),\;u_{4}= u(A_{x'},\;B_{y'},\;\lambda)$, and $a,\;b,\;c,\;d,\;e,\;f,\;g,\;h,\;i $  $\in$ $ \lbrace 1,-1 \rbrace $, with $0  \leq$ $ p $ $\leq$ 1.
From the above table we obtain, $ S = 4p+4$. Comparing this model with parts of Eq. \eqref{mm4} we have $ M_{1} = 2p $, which is consistent with \eqref{e12}. So, we have, $ F_{1} $ = $ 1- \frac{M_{1}}{2}$ = $ 1-\frac{2p}{2}=1-p $. Thus, to simulate GHZ bound ($ 4\sqrt{2} $) Alice offers 41$ \% $ freedom of choice in this scenario.

Meanwhile, this deterministic no-signaling model can simulate the general no-signaling bound \citep{52,53} of genuine tripartite non-locality by costing full freedom of measurement choice for Alice. This result holds good for other two parties settings also.

 \subsection{Relaxed $NS_{2}$ Inequality}\label{s3}
We have observed the effect of arbitrary one sided measurement dependence for Mermin and Svetlichny inequality in last two subsections. Now, we consider the $ NS_{2} $ inequality \cite{43} for the same, where  they showed that there exists correlations which violate this inequality (Eq. \eqref{e7}), but obey Svetlichny inequality. The $ NS_{2} $ inequality has the following form,
\begin{equation}\label{c5}
 \textit{R} = |\langle A_0B_0 \rangle +\langle A_0C_0 \rangle+\langle B_0C_1 \rangle-\langle A_1B_1C_0\rangle+\langle A_1B_1C_1\rangle|\leq 3; 
 \end{equation}

where  $\langle$$A_0$$B_0$$C_0$$\rangle$ = $\int{\rho(\lambda|xyz)\;u(A_x,\;\lambda)\;v(B_y,\;\lambda)\;w(C_z,\;\lambda)\;d\lambda}$; and\\
$\langle$$A_1$$B_1$$C_1$$\rangle$=$\int{\rho(\lambda|x',\;y',\;z')\;u(A_{x'},\;\lambda)\;v(B_{y'},\;\lambda)\;w(C_{z'},\;\lambda)\;d\lambda}$.\\ 

We define the quantity $S$ as, 
$S = \langle$$A_0$$B_0$$\rangle$+$\langle$$A_0$$C_0$$\rangle$$+$$\langle$$B_0$$C_1$$\rangle$$-$$\langle$$A_1$$B_1$$C_0$$\rangle$+$\langle$$A_1$$B_1$$C_1$$\rangle$.

For this local deterministic no-signaling model using the degree of measurement dependence $ M_{3} $ of Charlie's measurement settings, we have (for details, see Appendix C \eqref{sh3}),
\begin{equation}\label{c6}
 |S|\leq 3+M_{3}. 
 \end{equation}
For the state $ \vert GHZ \rangle =\dfrac{1}{\sqrt{2}} (\vert000\rangle  +\vert 111\rangle) $ with suitable measurement settings we find the value of of S = $ 1+2\sqrt{2} $ \cite{43}.

So, for the above model we have, $  3+M_{3} = 1+2\sqrt{2}$. Therefore, $ M_{3} = 2(\sqrt{2}-1)$, $ \Rightarrow F_{3}=0.59 $. 

Thus, this model can generate the GHZ state correlation with $59\% $ measurement freedom for Charlie's measurement setting. In this scenario, to generate the general no-signaling bound \citep{53}, we have $ M_{3} = 2 $, so Charlie will sacrifice his full freedom of choice to achieve this bound.

Now, rearranging the terms of S and envisaging from Alice's and Bob's point of view we found, 
\begin{eqnarray} \label{cc2}
 \vert S \vert \leq 3+ 2M_{1} + M_{3}\\
   \vert S \vert \leq 3+ 2M_{2} + M_{3}.
\end{eqnarray}
One can observe, Eq. \eqref{c6} is tight compared  to Eq. \eqref{cc2}.

Here we present a model which justify the above result;
\begin{table}  [h]
\caption{Deterministic no-signaling  model III}
\begin{tabular}{c c c c c c c c c c c c }
\hline
$\lambda$ & $u(A_x,\lambda)$&$u(A_{x'},\lambda)$&$v(B_y,\lambda)$&$v(B_{y'},\lambda)$& $w(C_z,\lambda)$& $w(C_{z'},\lambda)$&$\rho(\lambda|xy)$ &$\rho(\lambda|xz)$&$\rho(\lambda|yz')$&$\rho(\lambda|x'y'z)$&$\rho(\lambda|x'y'z')$\\ [1ex]
\hline
$\lambda_1$ &$ -a $&$a$&$-a$&$a$&$-a$&$-a$&1&1&1&0&0\\
$\lambda_2$ & $b$ &$b$&$b$&$b$&$b'$&$b'$&0&0&0&$1-p$&1\\
$\lambda_3$ & $c$ &$c$&$c$&$c$&$-c'$&$c'$&0&0&0&$p$&$0$\\[1ex]
\hline
\end{tabular}
\end{table}

Where $a,\; b,\; c $ $\in $ $ \lbrace 1,-1 \rbrace $, 0 $\leq$ $p$ $\leq$1, with $ b^{2} = b',\;c^{2} = c'$.
From the table we detect $S = 2p + 3$. So, $ M_3 = 2p $, signifies  $ F_3 $ = $1- \frac{M_{3}}{2}$ = $ 1-p $. Thus, to simulate GHZ state bound ($ 1 + 2 \sqrt{2} $) Charlie offers 41$ \% $ freedom of choice in this scenario.

Consequently, this deterministic model can generate the general no-signaling bound \citep{52,53} for $ p = 1$, by giving up full experimental freedom for Charlie.

\section{Bi-Party measurement dependence in Tripartite System}
We have already established the effect of arbitrary one party measurement relaxation for Mermin, Svetlichny and $ NS_{2} $ inequalities in the last section. Now, to proceed further, we want to find the effect of measurement dependence in randomly selected two parties measurement settings in tripartite scenario. We will use the quantities  $ M_{12},\; M_{23},\; M_{13} $ defined in Section \eqref{secII} in Eq. \eqref{e15}, for this purpose.
\subsection{Bipartite Relaxation of  Mermin Inequality }
In this section, we investigate the effect of arbitrary  measurement relaxations \cite{17,18,21} for any two parties measurement settings with the same tripartite measurement scenario for Mermin inequality. 

The Mermin inequality is, 
|$\langle$$A_1$$B_0$$C_0$$\rangle$+$\langle$$A_0$$B_1$$C_0$$\rangle$+$\langle$$A_0$$B_0$$C_1$$\rangle$-$\langle$$A_1$$B_1$$C_1$$\rangle$|\:$\leq$ 2.
We use the same notations as used in Section \eqref{s1}.

Here we have observed the measurement dependence effects of Alice-Bob, Bob-Charlie and Alice-Charlie measurement settings (for details, see Appendix D \eqref{sh4}), we finally have
 \begin{equation}\label{mm7}
\vert S\vert \leq 2+ Min \lbrace M_{12},\;M_{23},\;M_{13} \rbrace.
\end{equation}

This modified Mermin inequality is symmetric with respect to the parties, whereas we have observed an asymmetric relation in one-sided relaxed measurement settings.
In this scenario to simulate the GHZ state correlation, we assume $ 2+M_{12} = 4 $.
So, $M_{12} = 2;\;\Rightarrow F_{12} = 0 $. 

Thus, for arbitrary bipartite measurement dependence, the deterministic no-signaling model offering its full freedom of measurement choice to achieve the quantum bound of GHZ state, unlike one-sided measurement dependence case, where the same can be achieved with $50\%$ measurement dependence.

Now we present a model which is capable of exhibiting the above scenario.
\begin{table} [h]
\caption{Deterministic no-signaling  model IV}
\begin{tabular}{c c c c c c c c c c c}
\hline
$\lambda$ & $u(A_x,\lambda)$&$u(A_{x'},\lambda)$&$v(B_y,\lambda)$&$v(B_{y'},\lambda)$& $w(C_z,\lambda)$& $w(C_{z'},\lambda)$&$\rho(\lambda|x'yz)$ &$\rho(\lambda|xy'z)$&$\rho(\lambda|xyz')$&$\rho(\lambda|x'y'z')$\\ [1ex]
\hline
$\lambda_1$ & $a$ &$a$&$a$&$a$&$ a'$&$a'$&$1$&$1-p$&$1$&$1-p$\\
$\lambda_2$ &$ -b$ &$b$&$b$&$-b$&$b'$&$b'$&0&$p$&0&$p$\\[1ex]
\hline
\end{tabular}
\end{table}

Where $a,\; b$ $\in $ $\lbrace 1,-1\rbrace $, with $ a^{2}=a',\;b^{2}=b'  $ and $0 \leq$ $p$ $\leq$1.
Here, we have $ S = 2+2 p $. So, comparing  this model with the deterministic no-signaling model of Eq. \eqref{mm7}, we have $ M_{12} = 2 p $, which implies $ F_{12} = 1-p $. For $p = 1$, $ F_{12} = 0 $.

Thus this deterministic no-signaling model can generate the maximum quantum bound of Mermin inequality, S = 4 by costing its 100$\% $ measurement freedom for bipartite measurement settings.

\subsection{Bipartite Relaxation of Svetlichny Inequality }\label{s11}
In the last subsection, we have observed that arbitrary two-party measurement dependence deterministic model can generate the GHZ state correlation by sacrificing its full experimental free will in Mermin inequality. Here, we try to find the amount of measurement dependence is needed for the same cause in Svetlichny inequality\cite{6}. 

The required inequality is known as \cite{6}:

$R = |\langle$$A_0$$B_0$$C_0$$\rangle$$-$$\langle$$A_1$$B_1$$C_0$$\rangle$+$\langle$$A_0$$B_0$$C_1$$\rangle$$-$$\langle$$A_0$$B_1$$C_1$$\rangle$$+$$\langle$$A_1$$B_0$$C_0$$\rangle$+$\langle$$A_0$$B_1$$C_0$$\rangle$$-$$\langle$$A_1$$B_0$$C_1$$\rangle$$-$$\langle$$A_1$$B_1$$C_1$$\rangle| \leq 4$.

Assume $S = \langle$$A_0$$B_0$$C_0$$\rangle$+$\langle$$A_0$$B_0$$C_1$$\rangle$$+$$\langle$$A_1$$B_0$$C_0$$\rangle$$-$$\langle$$A_1$$B_0$$C_1$$\rangle$+$\langle$$A_0$$B_1$$C_0$$\rangle$$-$$\langle$$A_0$$B_1$$C_1$$\rangle$$-$$\langle$$A_1$$B_1$$C_0$$\rangle$$-$$\langle$$A_1$$B_1$$C_1$$\rangle$.

Here, we use the same notations as used in one-sided measurement dependence for Svetlichny inequality. We have the following simplification (see appendix E \eqref{sh5}),\\
\begin{equation}\label{pmm5}
\vert S\vert \leq 4 + Min \lbrace 4,\;2M_{12},\;2M_{23},\;2M_{13}\rbrace.
\end{equation}

For GHZ state the maximum quantum bound \cite{51} is S = $ 4\sqrt{2} $. So, in this arbitrary bipartite deterministic no-signaling model, we have $ 4+2M_{12} = 4\sqrt{2} $.
Therefore, $ M_{12} = 2(\sqrt{2}-1),\; \Rightarrow F_{12} = 0.59 $, i.e., for Svetlichny inequality arbitrary one-party and two-party measurement dependence models gives up the same amount of measurement freedom to reach maximum quantum value for GHZ state.

To simulate the general no-signaling bound \citep{53} for genuine tripartite non-locality, we have $ M_{12} = 2,\;\Rightarrow F_{12} = 0 $. Consequently, this model is also costing its full freedom of choice to generate the general no-signaling bound of Svetlichny inequality, likewise one-sided measurement dependence models.

A deterministic no-signaling model is given below (see TABLE V) in support of the above result.
 
   \begin{table} [h]
\caption{Deterministic no-signaling  model V}
\begin{tabular}{c c c c c c c c c c c c c c c}
\hline
$\lambda$ & $u_{1}$&$u_{2}$&$u_{3}$&$u_{4}$& $v(C_z,\lambda)$& $v(C_{z'},\lambda)$&$\rho(\lambda|xyz)$ &$\rho(\lambda|xyz')$&$\rho(\lambda|xy'z)$&$\rho(\lambda|xy'z')$&$\rho(\lambda|x'yz)$&$\rho(\lambda|x'yz')$&$\rho(\lambda|x'y'z)$&$\rho(\lambda|x'y'z')$\\ [1ex]
\hline
$\lambda_1$ & $a$ &$a$&$a$&$-a$&$a$&$a$&$1-p$&$1-p$&$1-p$&$1-p$&$1-p$&$1-p$&$1-p$&$1-p$\\
$\lambda_2$ & $b$ &$b$&$-b$&$b$&$b$&$b$&$p$&$p$&$0$&$0$&$0$&$0$&$0$&0\\
$\lambda_3$ & $c$ &$-c$&$c$&$-c$&$c$&$c$&0&0&$0$&$0$&$0$&0&$p$&$p$\\
$\lambda_4$ & $-d$ &$d$&$d$&$d$&$d$&$-d$&0&0&$p$&$p$&$0$&0&$0$&0\\
$\lambda_5$ & $e$ &$e$&$e$&$e$&$e$&$-e$&0&0&0&$0$&$p$&$p$&$0$&$0$\\[1ex]
\hline
\end{tabular}
\end{table}
Where $a,\; b,\; c,\; d,\;e $  $\in$ $\lbrace 1, -1 \rbrace $ and $0 \leq p \leq 1$.
Here, $ u_{1} = u(A_x,\;B_y,\;\lambda),\;u_{2} = u(A_x,\;B_{y'},\;\lambda),\;u_{3} = u(A_{x'},\;B_y,\;\lambda),\;u_{4} = u(A_{x'},\;B_{y'},\;\lambda) $.

From the  above table we have, S = $ 4+4p $, so comparing with Eq. \eqref{pmm5} we have $ M_{12} = 2p $, hence $ F_{12} = 1-p $. Thus, this model can generate the general no-signaling bound \citep{52,53} for $ p = 1$, which demonstrate 100$\% $ measurement dependence in bipartite  measurement settings.

\section{ Conclusion}
The assumption of measurement independence (freedom of choice) is very crucial  in deriving the Bell like  inequalities \cite{1,2}. In this present work we have derived, modified, relaxed bounds for Svetlichny, Mermin and NS$ _{2} $ inequalities with dichotomic measurement settings, respecting the assumptions of determinism and no signaling. We found that both single party and two-party (bipartite) measurement dependence models can violate the Bell like inequalities and it can simulate the quantum bound for GHZ state with $ 59\% $  measurement freedom in Svethlichny inequality, $NS_2$ inequality. Hence the degree measurement can be considered as a useful resource for simulating correlations which exhibit tripartite non-locality. While these models can achieve the general no-signaling bounds of their respective inequalities by sacrificing its full experimental free will. We also found model where it is offering $ 50\% $ free will to reach the quantum bound of GHZ state, i.e., Mermin inequality, while in two-party relaxation the respective parties cost its full freedom to generate quantum bound of GHZ state. Thus we can discriminate the genuine and standard tripartite non-locality with respect to measurement dependence. Thus our work will provide further insight to address this very idea of measurement dependence in multiparty non-local (more than three party) scenario \citep{54,55}, higher dimensional non-locality \citep{56,57} and other non-local scenarios \citep{58,59}.

{\bf Acknowledgment:}
S. H. acknowledge M. J. W. Hall for his valuable advice in forming this models. M. K. M. acknowledges G. Kar for  fruitful discussions. M. K. M. also acknowledges support from UGC, India, A. K. acknowledges support from CSIR, India and the authors I. Chattopadhyay and D. Sarkar acknowledge the work as part of QuEST initiatives by DST India.

\subsection*{APPENDIX A: RELAXED MERMIN INEQUALITY}\label{sh1}
In tripartite system with two measurement settings for each party and each measurement has two possible outcomes (as mentioned in Section \eqref{secI} ) standard tripartite non-locality is detected via the violation of  Mermin inequality \citep{42} which has the expression, 
 \begin{equation}\label{m1}
R= \vert\langle A_1B_0C_0 \rangle+\langle A_0B_1C_0 \rangle+\langle A_0B_0C_1 \rangle-\langle A_1B_1C_1 \rangle\vert\leq2.
\end{equation}
We consider S = $\langle$$A_1$$B_0$$C_0$$\rangle$ + $\langle$$A_0$$B_1$$C_0$$\rangle$ + $\langle$$A_0$$B_0$$C_1$$\rangle$ - $\langle$$A_1$$B_1$$C_1$$\rangle$. \\
Here in a deterministic no-signaling model, we apply arbitrary measurement dependence in either of the parties measurement settings. In this model the measurement outcomes are noted as  $ u(A_{x},\;\lambda),\;u(A_{x'},\;\lambda)  $ each taking values  from the set  $  \lbrace1, -1\rbrace $ for Alice's outcome. Identical interpretation follows for  $ v(B_{y},\;\lambda),\;v(B_{y'},\;\lambda) $ and  $ w(C_{z},\;\lambda),\;w(C_{z'},\;\lambda) $ accordingly.\\
 The correlation is defined as  $\langle$$A_1$$B_0$$C_0$$\rangle$ = $\int{\rho(\lambda|x^{'},\;y,\;z)u(A_{x'},\;\lambda)\; v(B_y,\;\lambda)\; w(C_z,\;\lambda)\;d\lambda}$, and similarly other terms of the expression follows as well. Using these quantities the parameter S then gives,\\
  S = $\int{\rho(\lambda|x',\;y,\;z)\;u(A_{x'},\;\lambda)\;v(B_y,\;\lambda)\;w(C_z,\;\lambda)\;d\lambda}$+ $\int{\rho(\lambda|x,\;y',\;z)\;u(A_x,\;\lambda)\;v(B_{y'},\;\lambda)\;w(C_z,\;\lambda)\;d\lambda}$\\ + $\int{\rho(\lambda|x,\;y,\;z')\;u(A_x,\;\lambda)\;v(B_y,\;\lambda)\;w(C_{z'},\;\lambda)\;d\lambda}$ $-$ $ \int{\rho(\lambda|x',\;y',\;z')\; u(A_{x'},\;\lambda)\; v(B_{y'},\;\lambda)\; w(C_{z'},\;\lambda)\;d\lambda}$.\\
  We next introduce the degree of measurement dependence factors in S by using simple mathematical process as follow,\\
     S = $\int{\rho(\lambda|x',\;y,\;z)\;u(A_{x'},\;\lambda)\;v(B_y,\;\lambda)\;w(C_z,\;\lambda)\;d\lambda}$$ - $$\int{\rho(\lambda|x,\;y,\;z)\;u(A_x,\;\lambda)\; v(B_y,\;\lambda)\; w(C_z,\;\lambda)\;d\lambda}$\\ + $\int{\rho(\lambda|x,\;y,\;z)\;u(A_x,\;\lambda)\;v(B_y,\;\lambda)\;w(C_z,\;\lambda)\;d\lambda}$ + $\int{\rho(\lambda|x,\;y',\;z)\; u(A_x,\;\lambda)\;v(B_{y'},\;\lambda)\;w(C_z,\;\lambda)\;d\lambda}$$\\ + $$\int{\rho(\lambda|x,\;y,\;z')\; u(A_x,\;\lambda)\;v(B_y,\;\lambda)\;w(C_{z'},\;\lambda)d\lambda}$ $-$ $\int{\rho(\lambda|x',\;y,\;z')\;u(A_{x'},\;\lambda)\;v(B_y,\;\lambda)\;w(C_{z'},\;\lambda)\;d\lambda}$\\ + $\int{\rho(\lambda|x',\;y,\;z')\;u(A_{x'},\;\lambda)\;v(B_y,\;\lambda)\;w(C_{z'},\;\lambda)\;d\lambda}$$ - $$\int{\rho(\lambda|x',\;y',\;z')\;u(A_{x'},\;\lambda)\;v(B_{y'},\;\lambda)\;w(C_{z'},\;\lambda)\;d\lambda}$.\\\\      
 = $\int{[\rho(\lambda|x',\;y,\;z)\;u(A_{x'},\;\lambda)-\rho(\lambda|x,\;y,\;z)\;u(A_x,\;\lambda)]\;v(B_y,\;\lambda)\;w(C_z,\;\lambda)\;d\lambda}$\\ + $\int{[\rho(\lambda|x,\;y,\;z)\;v(B_y,\;\lambda)+\rho(\lambda|x,\;y',\;z)\;v(B_{y'},\;\lambda)]\;u(A_x,\;\lambda)\;w(C_z,\;\lambda)\;d\lambda}$\\+$\int{[\rho(\lambda|x,\;y,\;z')\;u(A_x,\;\lambda) -\rho(\lambda|x',\;y,\;z')\;u(A_{x'},\;\lambda)]\;v(B_y,\;\lambda)\;w(C_{z'},\;\lambda)\;d\lambda}$\\+$\int{[\rho(\lambda|x',\;y,\;z')\;v(B_y,\;\lambda)-\rho(\lambda|x',\;y',\;z')\;v(B_{y'},\;\lambda)]\;u(A_{x'},\;\lambda)\;w(C_{z'},\;\lambda)\;d\lambda}$.\\\\
=$\int{[\rho(\lambda|x',\;y,\;z)-\rho(\lambda|x,\;y,\;z)\;\frac{u(A_x,\;\lambda)}{u(A_{x'},\;\lambda)}]\;u(A_x,\;\lambda)\;v(B_y,\;\lambda)\;w(C_z,\;\lambda)\;d\lambda}$\\+$\int{[\rho(\lambda|x,\;y,\;z)+\rho(\lambda|x,\;y',\;z)\;\frac{v(B_{y'},\;\lambda}{v(B_y,\;\lambda}]\;u(A_x,\;\lambda)\;v(B_y,\;\lambda)\;w(C_z,\;\lambda)\;d\lambda}$\\+$\int{[\rho(\lambda|x,\;y,\;z')-\rho(\lambda|x',\;y,\;z')\;\frac{u(A_{x'},\;\lambda)}{u(A_x,\;\lambda)}]\;u(A_x,\;\lambda)\;v(B_y,\;\lambda)\;w(C_{z'},\;\lambda)\;d\lambda}$\\+$\int{[\rho(\lambda|x',\;y,\;z')-\rho(\lambda|x',\;y',\;z')\;\frac{v(B_{y'},\;\lambda)}{v(B_y,\;\lambda)}]\;u(A_{x'},\;\lambda)\;v(B_y,\lambda)\;w(C_{z'},\;\lambda)\;d\lambda} .$\\
Let us assume, $ k_{1} = \frac{u(A_x,\;\lambda)}{u(A_{x'},\;\lambda)}$ and $ k_{2} = \frac{v(B_{y'},\;\lambda)}{v(B_y,\;\lambda)}$ then we have $ k_{1},\;k_{2} $ $ \in $ $ \lbrace 1, -1\rbrace $,
hence for $ k_{1}=1 $, we have
\begin{equation}\label{m2}
\vert S\vert \leq 2 + M_{2} + 2M_{1},
\end{equation}
and for $ k_{1} = -1,$ we have
\begin{equation}\label{m3}
\vert S\vert \leq 6+M_{2}.
\end{equation}
 We neglect this case as the model allow signaling \cite{40,52}. Whereas in other cases (Eq. \eqref{m2}) it violates Mermin inequality for non-zero values of $ M_{1}$ and $ M_{2} $. Following the same arguments and reordering the terms S gives,
\begin{equation}\label{m4}
\vert S \vert \leq 2 + 2 M_{1} + M_{3}.
\end{equation} 
Considering all the possible arrangement of terms and applying measurement dependence for the parties, we  finally have,
\begin{equation}
\vert S \vert\leq 2+  Min  \lbrace 2,\; 2M_{1} + M_{2},\;2M_{1} + M_{3},\;2M_{2} + M_{3},\;2M_{2} + M_{1},\;2M_{3} + M_{2},\;2M_{3} + M_{1} \rbrace.
\end{equation}
\subsection*{APPENDIX B: RELAXED SVETLICHNY INEQUALITY}\label{sh2}
Here our motivation is to establish the effect of  measurement dependence in one-sided measurement settings for Svetlichny inequality \citep{41}, which has the following representation,\\ 
\begin{equation}\label{c1}
R =\vert \langle A_0B_0C_0 \rangle +\langle A_0B_0C_1 \rangle+\langle A_1B_0C_0 \rangle-\langle A_1B_0C_1\rangle +\langle A_0 B_1C_0 \rangle - \langle A_0 B_1C_1 \rangle-\langle      A_1B_1C_0 \rangle-\langle A_1B_1C_1 \rangle \vert \leq 4. 
\end{equation}
Let us choose S = $\langle$$A_0$$B_0$$C_0$$\rangle$+$\langle$$A_0$$B_0$$C_1$$\rangle$$+$$\langle$$A_1$$B_0$$C_0$$\rangle$$-$$\langle$$A_1$$B_0$$C_1$$\rangle$+$\langle$$A_0$$B_1$$C_0$$\rangle$$-$$\langle$$A_0$$B_1$$C_1$$\rangle$$-$$\langle$$A_1$$B_1$$C_0$$\rangle$$-$$\langle$$A_1$$B_1$$C_1$$\rangle$. \\\\
 Where $\langle$ $A_0$$B_1$$C_0$ $\rangle$ = $\int{\rho(\lambda|x,\;y',\;z)\;u(A_x,\;B_{y'},\;\lambda)\;v(C_z,\;\lambda)\;d\lambda}$, with $-1\leq u(A_x,\;B_{y'},\;\lambda)\leq 1;\;-1\leq v(C_z,\lambda) \leq 1$,\\
 and similarly we define other terms of  S.\\
 Therefore\\
  S = $\int{\rho(\lambda|x,\;y,\;z)\;u(A_x,\;B_y,\;\lambda)\;v(C_z,\;\lambda)\;d\lambda}$+$\int{\rho(\lambda|x',\;y,\;z)\;u(A_{x'},\;B_y,\;\lambda)\;v(C_z,\;\lambda)\;d\lambda}$\\+$\int{\rho(\lambda|x,\;y,\;z')\;u(A_x,\;B_y,\;\lambda)\;v(C_{z'},\lambda)\;d\lambda}$ $-$ $\int{\rho(\lambda|x',\;y,\;z')\;u(A_{x'}\;B_y,\;\lambda)\;v(C_{z'},\;\lambda)\;d\lambda}$\\+$\int{\rho(\lambda|x,\;y',\;z)\;u(A_x,\;B_{y'},\;\lambda)\;v(C_z,\;\lambda)\;d\lambda}$$-$$\int{\rho(\lambda|x',\;y',\;z)\;u(A_{x'},\;B_{y'},\;\lambda)\;v(C_z,\;\lambda)\;d\lambda}$\\$-$ $\int{\rho(\lambda|x,\;y',\;z')\;u(A_x,\;B_{y'},\;\lambda)\;v(C_{z'},\;\lambda)\;d\lambda}$ $-$ $\int{\rho(\lambda|x',\;y',\;z')\;u(A_{x'},\;B_{y'},\;\lambda)\;v(C_{z'},\;\lambda)\;d\lambda}$.\\\\
 = $\int{u(A_x,\;B_y,\;\lambda)\;v(C_z,\;\lambda)\;[\rho(\lambda|x,\;y,\;z)+\rho(\lambda|x',\;y,\;z)\frac{u(A_{x'},\;B_y,\;\lambda)}{u(A_x,\;B_y,\;\lambda)}]\;d\lambda}$\\+$\int{u(A_x,\;B_y,\;\lambda)\;v(C_{z'},\lambda)\;[\rho(\lambda|x,\;y,\;z')-\rho(\lambda|x',\;y,\;z')\frac{u(A_{x'},\;B_y,\;\lambda)}{u(A_x,\;B_y,\;\lambda)}]\;d\lambda}$\\+$\int{u(A_x,\;B_{y'},\;\lambda)\;v(C_z,\;\lambda)\;[\rho(\lambda|x,\;y',\;z)-\rho(\lambda|x',\;y',\;z)\;\frac{u(A_{x'},\;B_{y'},\;\lambda)}{u(A_x,\;B_{y'},\;\lambda)}]\;d\lambda}$\\$-$ $\int{u(A_x,\;B_{y'},\;\lambda)\;v(C_{z'},\;\lambda)\;[\rho(\lambda|x,\;y',\;z')+\rho(\lambda|x',\;y',\;z')\;\frac{u(A_{x'},\;B_{y'},\;\lambda)}{u(A_x,\;B_{y'},\;\lambda)}]\;d\lambda}$.\\
$ \therefore $ $|S|$ $\leq$ $\int \vert{u(A_x,\;B_y,\;\lambda)\;v(C_z,\;\lambda)\vert \,\vert[\rho(\lambda|x,\;y,\;z)+\rho(\lambda|x',\;y,\;z)\;\frac{u(A_{x'},\;B_y,\;\lambda)}{u(A_x,\;B_y,\;\lambda)}]\vert \,d\lambda}$\\+$\int \vert{u(A_x,\;B_y,\;\lambda)\;v(C_{z'},\;\lambda)\vert \,\vert[\rho(\lambda|x,\;y,\;z')-\rho(\lambda|x',\;y,\;z')\;\frac{u(A_{x'},\;B_y,\;\lambda)}{u(A_x,\;B_y,\;\lambda)}]\vert \,d\lambda}$\\+$\int \vert{u(A_x,\;B_{y'},\;\lambda)\;v(C_z,\;\lambda)\vert \,\vert[\rho(\lambda|x,\;y',\;z)-\rho(\lambda|x',\;y',\;z)\;\frac{u(A_{x'},\;B_{y'},\;\lambda)}{u(A_x,\;B_{y'},\;\lambda)}]\vert \,d\lambda}$$\\+$$\int\vert{u(A_x,\;B_{y'},\;\lambda)\;v(C_{z'},\;\lambda)\vert\,\vert[\rho(\lambda|x,\;y',\;z')+\rho(\lambda|x',\;y',\;z')\;\frac{u(A_{x'},\;B_{y'},\;\lambda)}{u(A_x,\;B_{y'},\;\lambda)}]\vert \, d\lambda}$.\\\\
For deterministic model we have, $ u(A_x,\;B_y,\;\lambda) = \pm 1$, and $ v(C_z,\;\lambda) = \pm1, $ and similarly we select the values of other terms of S.\\
 Hence, $ \frac{u(A_{x'},\;B_y,\;\lambda)}{u(A_x,\;B_y,\;\lambda)}$ = $\pm1$ and so on. Consequently, taking $\frac{u(A_{x'},\;B_y,\;\lambda)}{u(A_x,\;B_y,\;\lambda)}$ = $\pm1$, and $\frac{u(A_{x'},\;B_{y'},\;\lambda)}{u(A_x,\;B_{y'},\;\lambda)}$ = $\pm1$, we have
\begin{equation}\label{c2}
|S| \leq 4+2 M_{1}. 
\end{equation}
Rearranging the terms and considering measurement dependence of Bob and Charlie setting it finally follows, 
\begin{equation}\label{c3}
\vert S\vert \leq 4+ Min\lbrace 4,\; 2M_{1},\;2M_{2},\;2M_{3} \rbrace.
\end{equation}
\subsection*{APPENDIX C: RELAXED NS$ _{2} $ INEQUALITY}\label{sh3}
Already we have seen the effect of arbitrary one sided measurement dependence for Mermin and Svetlichny inequality in last two subsections. Here we have drive modified $ NS_{2} $ inequality. The $ NS_{2} $ inequality \citep{43} has the following form,
\begin{equation}\label{c5}
 \textit{R} = |\langle A_0B_0 \rangle +\langle A_0C_0 \rangle+\langle B_0C_1 \rangle-\langle A_1B_1C_0\rangle+\langle A_1B_1C_1\rangle|\leq 3. 
 \end{equation}
Where  $\langle$$A_0$$B_0$$C_0$$\rangle$ = $\int{\rho(\lambda|x,\;y,\;z)\;u(A_x,\;\lambda)\;v(B_y,\;\lambda)\;w(C_z,\;\lambda)\;d\lambda}$.\\
$\langle$$A_1$$B_1$$C_1$$\rangle$ = $\int{\rho(\lambda|x',\;y',\;z')\;u(A_{x'},\;\lambda)\;v(B_{y'},\;\lambda)\;w(C_{z'},\;\lambda)\;d\lambda}$.\\ 
Let us suppose the quantity S as\\
S = $\langle$$A_0$$B_0$ $\rangle$ + $\langle$$A_0$$C_0$ $\rangle$$+$$\langle$$B_0$$C_1$ $\rangle$$-$$\langle$$A_1$$B_1$$C_0$$\rangle$+$\langle$$A_1$$B_1$$C_1$$\rangle$.\\
= $\int{\rho(\lambda|x,\;y)\;u(A_x,\;\lambda)\;v(B_y,\;\lambda)\;d\lambda}$+$\int{\rho(\lambda|x,\;z)\;u(A_x,\;\lambda)\;w(C_z,\;\lambda)\;d\lambda}$+$\int{\rho(\lambda|y,\;z')\;v(B_y,\;\lambda)\;w(C_{z'},\;\lambda)\;d\lambda}$\\$-$ $\int{\rho(\lambda|x',\;y',\;z)\;u(A_{x'},\;\lambda)\;v(B_{y'},\;\lambda)\;w(C_z,\;\lambda)\;d\lambda}$+$\int{\rho(\lambda|x',\;y',\;z')\;u(A_{x'},\;\lambda)\;v(B_{y'},\;\lambda)w(C_{z'},\;\lambda)\;d\lambda}$.\\\\
= $\int{\rho(\lambda|x,\;y)\;u(A_x,\;\lambda)\;v(B_y,\;\lambda)\;d\lambda}$+$\int{\rho(\lambda|x,\;z)\;u(A_x,\;\lambda)\;w(C_z,\;\lambda)\;d\lambda}$+$\int{\rho(\lambda|y,\;z')\;v(B_y,\;\lambda)\;w(C_{z'},\;\lambda)\;d\lambda}$\\$-$ $\int{\rho(\lambda|x',\;y',\;z)\;u(A_{x'},\;\lambda)\;v(B_{y'},\;\lambda)\;w(C_z,\;\lambda)\;d\lambda}$+$\int{\rho(\lambda\vert x',\;y',\;z')\;u(A_{x'},\;\lambda)\;v(B_{y'},\;\lambda)\;w(C_{z'},\lambda)\; d\lambda}$.\\\\\\
= $\int{\rho(\lambda|x,\;y)\;u(A_x,\;\lambda)\;v(B_y,\;\lambda)\;d\lambda}$+$\int{\rho(\lambda|x,\;z)\;u(A_x,\;\lambda)\;w(C_z,\;\lambda)\;d\lambda}$+$\int{\rho(\lambda|y,\;z')\;v(B_y,\;\lambda)\;w(C_{z'},\;\lambda)\;d\lambda}$\\+ $\int{[\rho(\lambda|x',\;y',\;z')-\rho(\lambda|x',\;y',\;z)\;\frac{w(C_z,\;\lambda)}{w(C_{z'},\;\lambda)}]\;u(A_{x'},\;\lambda)\;v(B_{y'},\;\lambda)\;w(C_{z'},\;\lambda)\;d\lambda}$.\\
For the deterministic no-signaling model using the degree of measurement dependence $ M_{3} $ of Charlie measurement settings, we have the following bound,
\begin{equation}
 |S|\leq 3+M_{3}. 
 \end{equation}
\subsection*{APPENDIX D: BI-PARTY RELAXATION OF MERMIN INEQUALITY}\label{sh4}
In this section we investigate the effect of arbitrary  measurement relaxation \cite{17,18,21} in any two-party measurement setting in the same tripartite measurement scenario for Mermin inequality\citep{42}.  \\
The Mermin inequality is \\
|$\langle$$A_1$$B_0$$C_0$$\rangle$+$\langle$$A_0$$B_1$$C_0$$\rangle$+$\langle$$A_0$$B_0$$C_1$$\rangle$-$\langle$$A_1$$B_1$$C_1$$\rangle$|\:$\leq$ 2.\\
We use the same notations as used in Section \eqref{s1}.\\
Therefore S = $\langle$$A_1$$B_0$$C_0$$\rangle$+$\langle$$A_0$$B_1$$C_0$$\rangle$+$\langle$$A_0$$B_0$$C_1$$\rangle$ $-$ $\langle$$A_1$$B_1$$C_1$$\rangle$.\\
 = $\int{\rho(\lambda|x',\;y,\;z)\;u(A_{x'},\;\lambda)\;v(B_y,\;\lambda)\;w(C_z,\;\lambda)\;d\lambda}$+$\int{\rho(\lambda|x,\;y',\;z)\;u(A_x,\;\lambda)\;v(B_{y'},\;\lambda)\;w(C_z,\;\lambda)\;d\lambda}$\\+$\int{\rho(\lambda|x,\;y,\;z')\;u(A_x,\;\lambda)\;v(B_y,\;\lambda)\;w(C_{z'},\;\lambda)\;d\lambda}$$-$$\int{\rho(\lambda|x',\;y',\;z')\;u(A_{x'},\;\lambda)\;v(B_{y'},\;\lambda)\;w(C_{z'},\;\lambda)\;d\lambda}$.\\
= $ \int d\lambda\; [\rho(\lambda \vert x,\;y,\;z')$$-$$\rho(\lambda \vert x',\;y',\;z')\; \frac{u(A_{x'},\;\lambda)\;v(B_{y'},\;\lambda)}{u(A_x,\;\lambda)\;v(B_y,\;\lambda)}]\; u(A_x,\;\lambda)\;v(B_y,\;\lambda)\;w(C_{z'},\;\lambda)$\\
+$\int d\lambda \;[\rho(\lambda \vert x,\;y',\;z)$+$\rho(\lambda \vert x',\;y,\;z) \;\frac{u(A_{x'},\;\lambda)\;v(B_y,\;\lambda)}{u(A_x,\;\lambda)\;v(B_{y'},\lambda)}]\; u(A_x,\;\lambda)\;v(B_{y'},\;\lambda)\;w(C_z,\;\lambda)$.\\\\
Let us define $ k_1 = \frac{u(A_{x'},\;\lambda)\;v(B_{y'},\;\lambda)}{u(A_x,\;\lambda)\;v(B_y,\;\lambda)}, \; k_2 = \frac{u(A_{x'},\;\lambda)\;v(B_y,\;\lambda)}{u(A_x,\;\lambda)\;v(B_{y'},\;\lambda)}$.\\
Then for a deterministic no-signaling model we must have, $ k_1,\;k_2 \in \lbrace 1,-1\rbrace $.\\ 
Therefore for all possible values of $ k_1,\;k_2, $ we found
\begin{eqnarray} \label{cc3}
\vert S\vert \leq 2+ M_{12}.
\end{eqnarray}
 Here rearranging the terms and considering the effect of measurement dependence for Bob-Charlie and Alice-Charlie measurement settings, we finally have the modified inequality,
 \begin{equation}\label{p1}
\vert S\vert \leq 2+ Min \lbrace M_{12},\; M_{23},\; M_{13} \rbrace.
\end{equation}
\subsection*{APPENDIX E: BI-PARTY RELAXATION OF SVETLICHNY INEQUALITY}\label{sh5}
In this last appendix we have founded the modified Svetlichny inequality, when any two parties have restricted measurement settings in a tripartite non-locality scenario. The required inequality is known as \citep{41}\\
R = |$\langle$$A_0$$B_0$$C_0$$\rangle$$-$$\langle$$A_1$$B_1$$C_0$$\rangle$+$\langle$$A_0$$B_0$$C_1$$\rangle$$-$$\langle$$A_0$$B_1$$C_1$$\rangle$$+$$\langle$$A_1$$B_0$$C_0$$\rangle$+$\langle$$A_0$$B_1$$C_0$$\rangle$$-$$\langle$$A_1$$B_0$$C_1$$\rangle$$-$$\langle$$A_1$$B_1$$C_1$$\rangle$| $\leq$ 4.\\\\
Let us assume S = $\langle$$A_0$$B_0$$C_0$$\rangle$+$\langle$$A_0$$B_0$$C_1$$\rangle$$+$$\langle$$A_1$$B_0$$C_0$$\rangle$$-$$\langle$$A_1$$B_0$$C_1$$\rangle$+$\langle$$A_0$$B_1$$C_0$$\rangle$$-$$\langle$$A_0$$B_1$$C_1$$\rangle$$-$$\langle$$A_1$$B_1$$C_0$$\rangle$$-$$\langle$$A_1$$B_1$$C_1$$\rangle$.\\
Here we use the same notations as used in one-sided measurement dependence for Svetlichny inequality, we have the following simplification,\\
S = $\int{\rho(\lambda|x,\;y,\;z)\;u(A_x,\;B_y,\;\lambda)\;v(C_z,\;\lambda)\;d\lambda}$$-$$\int{\rho(\lambda|x',\;y',\;z)\;u(A_{x'},\;B_{y'},\;\lambda)\;v(C_z,\;\lambda)\;d\lambda}$\\+$\int{\rho(\lambda|x,\;y,\;z')\;u(A_x,\;B_y,\;\lambda)\;v(C_{z'},\;\lambda)\;d\lambda}$ $-$ $\int{\rho(\lambda|x,\;y',\;z')\;u(A_x,\;B_{y'},\;\lambda)\;v(C_{z'},\;\lambda)\;d\lambda}$\\+$\int{\rho(\lambda|x',\;y,\;z)\;u(A_{x'},\;B_y,\;\lambda)\;v(C_z,\;\lambda)\;d\lambda}$+$\int{\rho(\lambda|x,\;y',\;z)\;u(A_x,\;B_{y'},\;\lambda)\;v(C_z,\;\lambda)\;d\lambda}$\\$-$ $\int{\rho(\lambda|x',\;y,\;z')\;u(A_{x'},\;B_y,\;\lambda)\;v(C_{z'},\;\lambda)\;d\lambda}$$-$$\int{\rho(\lambda|x',\;y',\;z')\;u(A_{x'},\;B_{y'},\;\lambda)\;v(C_{z'},\;\lambda)\;d\lambda}$.\\\\
 = $ \int d\lambda \;[\rho (\lambda\vert x,\;y,\;z)$$-$$\rho (\lambda\vert x',\;y',\;z) \;\frac{u(A_{x'},\;B_{y'},\;\lambda)}{u(A_x,\;B_y,\;\lambda)}]\; u(A_x,\;B_y,\;\lambda)\;v(C_z,\;\lambda) $\\+$ \int d\lambda\; [\rho (\lambda\vert x,\;y,\;z')$$-$$\rho (\lambda\vert x',\;y',\;z')\; \frac{u(A_{x'},\;B_{y'},\;\lambda)}{u(A_x,\;B_y,\;\lambda)}] u(A_x,\;B_y,\;\lambda)\;v(C_{z'},\;\lambda) $\\ $-$ $ \int d\lambda\; [\rho (\lambda\vert x,\;y',\;z')$+$\rho (\lambda\vert x',\;y,\;z') \;\frac{u(A_{x'},\;B_y,\;\lambda)}{u(A_x,\;B_{y'},\;\lambda)}] \;u(A_x,\;B_{y'},\;\lambda)\;v(C_{z'},\;\lambda) $\\+$ \int d\lambda\; [\rho (\lambda\vert x',\;y,\;z)$+$\rho (\lambda\vert x,\;y',\;z)\; \frac{u(A_x,\;B_{y'},\;\lambda)}{u(A_{x'},\;B_y,\;\lambda)}]\; u(A_{x'},\;B_y,\;\lambda)\;v(C_z,\;\lambda) $.\\
$\therefore $ Considering $ k_1 = \frac{u(A_x,\;B_y,\;\lambda)}{u(A_{x'},\;B_{y'},\;\lambda)}, \; k_2 = \frac{u(A_{x'},\;B_y,\;\lambda)}{u(A_x,\;B_{y'},\;\lambda)}$.\\
In a deterministic no-signaling model we must have, $ k_1,\; k_2 \in \lbrace 1,-1\rbrace $, therefore S reduced to the following, \\
\begin{equation}\label{p4}
\vert S\vert\leq 4+2 M_{12}.
\end{equation}
By rearranging the terms we have two similar  quantities for Alice-Charlie and Bob-charlie restricted settings. Combining all the possible results we finally have, 
\begin{equation}
\vert S\vert \leq 4 + Min \lbrace 4,\;2M_{12},\;2M_{23},\;2M_{13}\rbrace.
\end{equation}

\end{document}